\documentclass[10pt,twocolumn,aps,english,pra,superscriptaddress,showpacs]{revtex4}
\usepackage{amsfonts}
\usepackage[T1]{fontenc}
\usepackage[latin9]{inputenc}
\usepackage{amsmath}
\usepackage{amssymb}
\usepackage{graphicx}
\usepackage{babel}
\usepackage{booktabs}
\usepackage{tabu}

\makeatletter
\@ifundefined{textcolor}{}
{
 \definecolor{BLACK}{gray}{0}
 \definecolor{WHITE}{gray}{1}
 \definecolor{RED}{rgb}{1,0,0}
 \definecolor{GREEN}{rgb}{0,1,0}
 \definecolor{BLUE}{rgb}{0,0,1}
 \definecolor{CYAN}{cmyk}{1,0,0,0}
 \definecolor{MAGENTA}{cmyk}{0,1,0,0}
 \definecolor{YELLOW}{cmyk}{0,0,1,0}
 }
\makeatother

\begin{document}

\title{The Coupling-Deformed Pointer Observables and Weak Values}
\author{Yu-Xiang Zhang}
\affiliation{Hefei National Laboratory for Physical Sciences at Microscale, Department of
Modern Physics, University of Science and Technology of China, Hefei, Anhui
230026, China}
\affiliation{The CAS Center for Excellence in QIQP and the Synergetic Innovation Center
for QIQP, University of Science and Technology of China, Hefei, Anhui
230026, China}
\affiliation{Kuang Yaming Honors School, Nanjing University, Nanjing, Jiangsu 210023,
China}
\author{Shengjun Wu}
\email{sjwu@nju.edu.cn}
\affiliation{Kuang Yaming Honors School, Nanjing University, Nanjing, Jiangsu 210023,
China}
\author{Zeng-Bing Chen}
\email{zbchen@ustc.edu.cn}
\affiliation{Hefei National Laboratory for Physical Sciences at Microscale, Department of
Modern Physics, University of Science and Technology of China, Hefei, Anhui
230026, China}
\affiliation{The CAS Center for Excellence in QIQP and the Synergetic Innovation Center
for QIQP, University of Science and Technology of China, Hefei, Anhui
230026, China}
\date{\today }

\begin{abstract}
While the novel applications of weak values have recently attracted wide
attention, weak measurement, the usual way to extract weak values, suffers
from risky approximations and severe quantum noises. In this paper, we show
the weak-value information can be obtained exactly in strong measurement
with post-selections, via measuring the coupling-deformed pointer
observables, i.e., the observables selected according to the coupling
strength. With this approach, we keep all the advantages claimed by weak-measurement 
schemes and at the same time solve some widely criticized
problems thereof, such as the questionable universality, systematical bias,
and drastic inefficiency.
\end{abstract}

\maketitle

\section{Introduction}

Introduced by Aharonov, Albert and Vaidman (AAV) \cite{weak
AAV,vaidman,today} about thirty years ago, the \emph{weak value} arises in
the outcome of weak measurement with postselection,
which is conventionally abbreviated as \emph{weak measurement}.
Weak values play the same role in weak measurements as expectation values play in von Neumann measurements.
The central task of
many impressive applications of weak measurements is then to determine the weak values which could
incorporate the desired information \cite{review}.
For example, weak-value tomography, to which we shall pay more attention below, has realized the direct
measurements of both quantum states \cite{wave
function,27-dimension,jp,Dirac,wu,Dirac distribution,Dirac-2,photon
direct,comwave} and quantum dynamical processes \cite{xiang}. The directness
and simplicity of weak-value tomography in implementation make it perhaps
the unique option for the reconstructions of high-dimensional states, and
the current record is $19200$-dimensional states \cite{comwave}. Such kind
of achievement is practically impossible for standard tomography techniques
\cite{book}.

We shall briefly comment on how weak measurements perform better for many cases in the next section.
Nevertheless, some severe disadvantages of weak-measurement
applications have been exposed. A common
trouble arises from the possible failure of weak value as the real pointer
reading, especially when the weak value approaches infinity \cite%
{wu-li,pang,loren,zhu-zhang}. It makes weak-measurement tomography not
universal \cite{review,validity,simulation}. This shortcoming also limits
the performance of weak-value amplification \cite{science,review}. Moreover,
weak measurements suffer from serious quantum noise, which is usually argued as the inevitable price.
For example, to suppress statistical error down to the same level,
weak measurements require several orders of magnitude more samples than the
standard tomography scheme \cite{simulation}. Besides that, the weak limit is
inaccessible since only finite (even though tiny) coupling strength can be
used in experiments. Systematical errors are then introduced unavoidably and
behave as inevitable bias in weak-value tomography \cite{simulation}.

Instead of obtaining the weak-value information approximately as in the
originally proposed measurement at the weak limit, in this article, we show
that one can get it exactly in the original setup but
with interaction of arbitrary strength, provided that
the observable read on the pointer is properly chosen as a coupling-deformed
(CD) pointer observable. This is our main idea.
Our method works for the whole regime of measurement strength. The exactness and the
utilization of stronger measurements remove the problems of risky approximations
and quantum noise in the determinations of weak values.
Particularly, in the weak-value
tomography, our method requires only a slight alteration to the
current experimental setting and thus can be realized in
experiments straightforwardly.

We arrange this paper as followed. In Sec. \ref{section-aav formalism},
we shall introduce the standard formulas of weak values and weak measurements,
and the application in quantum state tomography. In Sec. \ref{main section}
we give our main results on coupling-deformed pointer observables
and demonstrate how our method solves the mentioned problems.
In Sec. \ref{section poor} we propose methods against possible
complexities arising in the method of CD observables. Finally we discuss further
implications and conclude this article in Sec. \ref{section discussion}.

\section{Aharonov, Albert, and Vaidman's formalism of weak value}\label{section-aav formalism}

In this section, we shall briefly introduce AAV's
perturbation formalism of weak measurements, and the applications
of weak value in quantum state tomography.

\subsection{Weak measurement and weak value}
The model in consideration consists of a system to be
measured and a pointer.
Suppose the system is initialized in the state $\rho
_{in}=|\psi _{in}\rangle \langle \psi _{in}|$. To measure observable $\hat{A}
$, one couples the system to the pointer initialized in the state $|\phi
_{0}\rangle $, via the typical unitary time evolution operator $U=\exp (-ig%
\hat{A}\otimes \hat{p})$, where $\hat{p}$ is defined on the pointer and $g$
is the coupling strength (assumed to be dimensionless).
After that, the system is postselected into $\Pi _{f}=|\psi _{f}\rangle \langle \psi _{f}|$.
If the coupling is weak enough, the unnormalized pointer state $\langle \psi _{f}|U|\psi
_{in}\rangle |\phi _{0}\rangle $ will be \cite{weak AAV,vaidman}
\begin{equation}
\begin{aligned}
\langle\psi_f|(I-ig\hat{A}&\otimes\hat{p})|\psi_{in}\rangle|\phi_0%
\rangle+O(g^2)\\
\approx&\langle\psi_f|\psi_{in}\rangle\exp(-igA_w\hat{p})|\phi_0\rangle,\;\;%
\;g\rightarrow 0. \end{aligned}  \label{AAVf}
\end{equation}%
This rough derivation suggests that if $\hat{p}$ is the momentum operator
and $\hat{q}$ is the position operator ($[\hat{q},\hat{p}]=i\hbar $), then
pointer's $\hat{q}$-reading will be shifted by $gA_{w}$. Here
\begin{equation}
A_{w}=\frac{\langle \psi _{f}|\hat{A}|\psi _{in}\rangle }{\langle \psi
_{f}|\psi _{in}\rangle }=\frac{\mathrm{tr}(\Pi _{f}\hat{A}\rho _{in})}{%
\mathrm{tr}(\Pi _{f}\rho _{in})}  \label{weakvalue}
\end{equation}%
is known as the \emph{weak value}. One can determine the real and imaginary
parts of weak value, respectively, by measuring two pointer observables,
which are usually denoted by $\hat{q}$ and $\hat{p}$ \cite{weak
AAV,vaidman,jozsa}. The outcomes of such measurements,
or the readings of the pointer, are denoted by $\langle \hat{q}%
\rangle _{f}$ and $\langle \hat{p}\rangle _{f}$, respectively.
The subscript ``f'' denotes that the expectation values are
conditioned on successful postselection.
The postselection is equivalent to a projective measurement.
Thus the action of postselection and reading the pointer is equivalent to
a measurement of $\Pi_f\otimes\hat{q}(\hat{p})$ on the entire system.
It implies that
$$ \langle\hat{q}(\hat{p})\rangle_f=\frac{\langle\Pi_f\otimes\hat{q}(\hat{p})\rangle}{P_f}$$
where on the right hand side $\langle \cdot \rangle $ stands for the
average value in the coupled joint state $U|\psi _{in}\rangle |\phi
_{0}\rangle $; $P_f=\langle\Pi_f\otimes I\rangle$ ($I$ is the identity operator) is
the probability of successful postselection.
In the weak limit $g\rightarrow 0$, $P_f$ will equal the
denominator of $A_w$, $\mathrm{tr}(\Pi_f\rho_{in})$.

\subsection{Applications}

Here we investigate applications to determine the unknown information
incorporated in the weak value. These applications are classified into
two categories, ``measurable complex values'' and ``conditioned average'',
in Ref. \cite{review}. From the definition of weak value (\ref{weakvalue}),
it is obvious that the numerator $\mathrm{tr}(\Pi _{f}\hat{A}\rho _{in})$ is what we are really interested in,
since the denominator is directly accessible in the projective measurements.

As the natural approach to acquire weak values, weak measurements with postselections
are employed in these applications. And in fact, these applications are different
in the specific selections of
$|\psi_{in}\rangle$, $|\psi_f\rangle$ and $\hat{A}$.
Thus we shall not elaborate on them but merely focus on weak-value tomography,
or direct measurement of quantum states.

\subsubsection{Weak-value tomography}

The numerator $\mathrm{tr}(\Pi_f\hat{A}\rho_{in})$ in the definition of $A_w$
is actually what one wants in weak-value tomography, since it gives the wave
functions \cite{27-dimension,wave function,comwave,jp} or the Kirkwood-Dirac
distribution of general states \cite{wu,Dirac,Dirac
distribution,Dirac-2,photon direct}. To see it, if $\hat{A}=|x\rangle
\langle x|$ (the projector at position \emph{x}) and $|\psi _{f}\rangle $ is
the zero-momentum eigenstate $|\vec{0}_{p}\rangle$, then
\begin{equation}
(|x\rangle\langle x|)_w=\frac{\psi _{in}(x)\langle\psi_{in}|\vec{0}%
_{p}\rangle}{|\langle\vec{0}_{p}|\psi_{in}\rangle|^2}
\end{equation}
where we have assumed the normalization $\langle x|\vec{0}_p\rangle=1$. Thus, it
is the numerator that gives the wave function $\psi _{in}(x)$ \cite{wave
function}.

In the experiment by Lundeen \textit{et al}. \cite{wave function}, the
pointer is played by photon's polarization initialized in $|0\rangle $ ($%
\sigma _{z}|0\rangle =|0\rangle $). To get the weak-value information from
measurements, the pointer observables, $\hat{p}$ and $\hat{q}$, are
substituted with two Pauli operators, $\hat{p}\rightarrow \sigma _{x}$ and $%
\hat{q}\rightarrow \sigma _{y}$. Explicitly, the formula for the numerator
of $(|x\rangle \langle x|)_{w}$ is
\begin{equation}
\psi _{in}(x)\langle \psi _{in}|\vec{0}_{p}\rangle \approx \frac{-1}{2g}%
\langle \Pi _{f}\otimes \hat{q}\rangle +\frac{i}{2g}\langle \Pi _{f}\otimes
\hat{p}\rangle .  \label{waveqp}
\end{equation}
This equation also shows how the unknown wave function (left-hand side)
is determined from the outcome of measurements (right-hand side).
The formula for mixed states goes in similar ways \cite{wu,Dirac distribution}.

\subsubsection{Merits and drawbacks}

The merits of weak-value tomography were highlighted in
all the relevant references \cite{27-dimension,wave function,comwave,jp,wu,Dirac,Dirac
distribution,Dirac-2,photon direct}.
In sharply contrast to the global inversion required in the standard tomography,
the most striking feature of weak-value tomography is
exhibited in the direct extraction of the wave function at each spatial point.
The entire wave function can be generated in real time.
The second merit is the simplicity of manipulation.
The observable $\hat{A}$ is chosen from the set $\{|x\rangle\langle x|\}_x$.
These projectors commute with each other, and correspond to one orthogonal basis of the Hilbert space.
To compare, the observables required in the standard tomography do not mutually commute.
In the cases in which $\rho_{in}$ is a mixed state, the postselection should be extended
to $\{\Pi_f\}_f$ ($\Pi_f\Pi_{f'}=\Pi_f\delta_{f,f'}$, $\sum_f\Pi_f=I$, $I$ is the identity operator).
These postselections in weak-value tomography are compatible so that can be realized a single experimental setup.
Because of these factors, weak-value tomography was successfully applied in the estimation of high-dimension states,
like 27 dimensional states in \cite{27-dimension} and 192 dimensional states in \cite{comwave}.

One may wonder that why we do not directly measure the observable $\Re(\Pi_f\hat{A})$ and
$\Im(\Pi_f\hat{A})$. If the projective measurements of the two observables are available in
the laboratory, perhaps that would be better.
However, here they are written in the form of $|x\rangle\langle\vec{0}_p|+|\vec{0}_p\rangle\langle x|$
and $i|x\rangle\langle\vec{0}_p|-i|\vec{0}_p\rangle\langle x|$.
It seems impossible to directly implement the measurements of them with the current capability.
Moreover, in the context of tomography, measuring the complete set of these observables is
just another version of the standard tomography.

The drawbacks of weak-measurement schemes come from the approximations and the weakness.
The error of the approximations used in Eq. (\ref{AAVf}) is drastic when $\mathrm{tr}%
(\Pi_f\rho_{in})$ is close to zero \cite{wu-li,pang,loren}. To avoid this
trouble one should know enough information about $\rho_{in}$, which seems
impossible since $\rho_{in}$ is preassumed as unknown. It
makes the method not universal \cite{review,validity,simulation}. Other
problems, such as bias and inefficiency, have been mentioned above.
These drawbacks are rooted in the original formalism of weak measurements and weak value and 
thus cannot be solved trivially.

\section{Coupling-Deformed Pointer Observable}\label{main section}

\begin{figure}[htb]
    \includegraphics[width=0.4\textwidth]{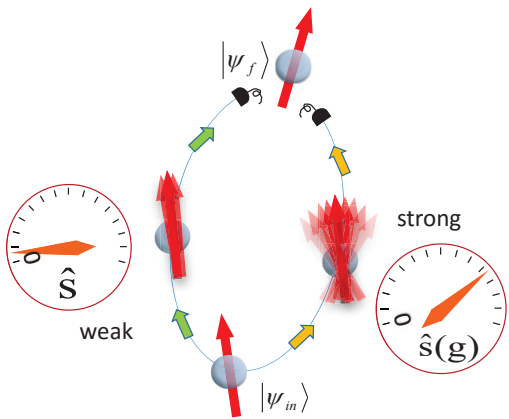}
    \caption{The extraction of weak-value information starts from
    the initial state preparation of $|\psi_{in}\rangle$, and ends at the
    postselection of the system onto $|\psi_f\rangle$. AAV's formalism requires
    the measurement to be as weak as possible (left path in the figure). While we
    show that we can get the weak-value information using measurements at an arbitrary strength, say
    $g$, if we replace the pointer observable with the coupling-deformed pointer observable
    $\hat{s}(g)$
    (right path).}
    \label{fig-1}
\end{figure}

In this section, we will introduce the concept of
the coupling-deformed pointer observable, which
permits us to extract the weak-value information exactly with measurements
of any strength. See Fig. \ref{fig-1}
for an overview.

\subsection{A nonperturbative framework}
To go beyond the first-order perturbation \cite%
{wu-li,loren,non-pur,gen-von,zhu-zhang,contex,contex-1}, the popular
method is to keep more terms of the series expansion
\begin{equation}
U=\sum_{n=0}^{\infty }\frac{(-ig)^{n}}{n!}\hat{A}^{n}\otimes \hat{p}^{n}.
\label{expansion}
\end{equation}%
However, expressions written with a long summation are too complicate to
be used for further analysis and applications. For our purpose, a simple and
fast nonperturbative approach is posteriorly proved to be much better.

Suppose $\hat{A}=\sum_{m=1}a_{m}|a_{m}\rangle \langle a_{m}|$ with $\langle
a_{m}|a_{n}\rangle =\delta _{mn}$ and $a_m\neq 0$. Then one has $|\psi _{in}\rangle
=\sum_{m=1}c_{m}|a_{m}\rangle +c_{0}|\psi ^{\bot }\rangle $ where $|\psi
^{\bot }\rangle $ is the component of $|\psi _{in}\rangle $ in the null space
of $\hat{A}$. After the coupling $U=\exp (-ig\hat{A}%
\otimes \hat{p})$, the overall state becomes
	\begin{equation}
	c_{0}|\psi ^{\bot }\rangle |\phi _{0}\rangle +\sum_{m=1}c_{m}|a_{m}\rangle
	|\phi _{m}(g)\rangle ,  \label{combined-state}
	\end{equation}%
where $|\phi _{m}(g)\rangle =\exp (-iga_{m}\hat{p})|\phi _{0}\rangle $. We
denote $|\psi ^{\bot }\rangle $ as $|a_{0}\rangle $ for convenience. Suppose
$\hat{s}\in \{\hat{p},\hat{q}\}$, momentum or position when the pointer is a continuous-variable system,
or Pauli operators when the pointer is a qubit system,
or other observables of the pointer accordingly. Then the expectation value
of $\Pi _{f}\otimes \hat{s}$ for the given coupling strength $g$ can be
written in a compact form as
\begin{subequations}
\label{subequation}
\begin{align}
& \langle \Pi _{f}\otimes \hat{s}\rangle _{g}=g\sum_{m,n=0}(\rho
_{in})_{mn}(\Pi _{f})_{nm}Q_{nm}(g,\hat{s}),  \label{key} \\
& Q_{nm}(g,\hat{s})=\frac{1}{g}\langle \phi _{n}(g)|\hat{s}|\phi
_{m}(g)\rangle .
\end{align}%
where $(\rho _{in})_{mn}=\langle a_{m}|\rho _{in}|a_{n}\rangle $, $(\Pi
_{f})_{nm}=\langle a_{n}|\Pi _{f}|a_{m}\rangle $. Additionally,
when the input state $\rho_{in}$ is not pure, the definition of
$|a_0\rangle$ becomes ambiguous. In this case, suppose $%
|\psi _{f}\rangle =\sum_{m=1}\beta _{m}|a_{m}\rangle +\beta _{0}|\psi
_{f}^{\bot }\rangle $. Then $|a_{0}\rangle $ could be redefined as $|\psi
_{f}^{\bot }\rangle $. It is not complicated to check that this redefinition
will not change the value of the right-hand side of Eq. (\ref{key}),
which then turns out to be applicable to
situations with $\rho _{in}$ being a mixed state.

\subsection{A stronger sufficient condition for the extraction of weak-value information}
Within the above nonperturbation framework, let us investigate how
the numerator of the weak value, $\mathrm{tr}(\Pi _{f}\hat{A}\rho _{in})$,
arises at the weak limit. The result will hint at a sufficient condition
for the extraction of weak-value information that is stronger than the weak limit.

Under the condition that $\langle \phi _{0}|\hat{s}|\phi _{0}\rangle =0$,
the $Q$-matrix elements read in the weak limit as
\end{subequations}
\begin{equation}
\begin{aligned} \lim_{g\rightarrow 0}Q_{nm}(g,\hat{s})&=ia_{n}\langle \phi
_{0}|\hat{p}\hat{s}|\phi _{0}\rangle -ia_{m}\langle \phi
_{0}|\hat{s}\hat{p}|\phi _{0}\rangle\\ &\equiv Q_{nm}(0,\hat{s})
\label{order-1} \end{aligned}
\end{equation}%
Substituting Eq. (\ref{order-1}) into Eq. (\ref{key}), one gets
\begin{equation}
\begin{aligned} \langle\Pi_f\otimes\hat{s}\rangle_{g\rightarrow0}=
-ig\langle\phi_0|[\hat{s},\hat{p}]_{-}|\phi_0\rangle\,
\mathfrak{Re}\,[\mathrm{tr}(\Pi_f\hat{A}\rho_{in})]\\
+g\langle\phi_0|[\hat{s},\hat{p}]_+|\phi_0\rangle\,
\mathfrak{Im}\,[\mathrm{tr}(\Pi_f\hat{A}\rho_{in})] \end{aligned}
\label{weak-value-order1}
\end{equation}%
Here $\mathfrak{Re}$ and $\mathfrak{Im}$ mean the real and imaginary parts,
respectively, and $[\hat{s},\hat{p}]_{\pm }=\hat{s}\hat{p}\pm \hat{p}\hat{s}$.
That means, for the real part of weak-value numerator, one should read a
point observable $\hat{q}$, i.e., $\hat{s}\rightarrow \hat{q}$, which
satisfies $\langle \phi _{0}|[\hat{q},\hat{p}]_{+}|\phi _{0}\rangle =0$, and
for the imaginary part, one can read $\hat{s}\rightarrow \hat{p}$, which
obviously satisfies $\langle \phi _{0}|[\hat{p},\hat{p}]_{-}|\phi
_{0}\rangle =0$. Explicitly,
\begin{equation}
\begin{aligned} &\mathfrak{Re}\,[\mathrm{tr}(\Pi_f\hat{A}\rho_{in})]=
\frac{i}{g\langle\phi_0|[\hat{q},\hat{p}]_{-}|\phi_0\rangle}\langle\Pi_f%
\otimes\hat{q}\rangle_{g\rightarrow 0},\\
&\mathfrak{Im}\,[\mathrm{tr}(\Pi_f\hat{A}\rho_{in})]=
\frac{1}{g\langle\phi_0|[\hat{p},\hat{p}]_+|\phi_0\rangle}\langle\Pi_f%
\otimes\hat{p}\rangle_{g\rightarrow 0}. \end{aligned}  \label{real-ima-weak}
\end{equation}%
These two expressions verify the sufficiency of weak limit,
while the derivation for Eq. (\ref{real-ima-weak}) illuminates a wider
sufficient condition:
the elements of the $Q$ matrix, $Q(g,\hat{s})$, have the special type that
$Q_{mn}=c a_m+c^* a_n$ with $c$ being a constant. To reach this type of
$Q$ matrix, weak limit is sufficient, but not necessary.

\subsection{Coupling-deformed pointer observables}

In the weak limit,
Eqs. (\ref{order-1}) and (\ref{weak-value-order1}) suggest that
the specific selection of $\hat{s}$ is not important: we just need to make sure that
the factor $\langle\phi_0|\hat{p}\hat{s}|\phi_0\rangle$ is real or imaginary,
while to go beyond the weak regime, the key observation is
that we can exploit the untapped freedom of choosing a proper $\hat{s}$.

Let us denote the modified pointer observable as $\hat{s}(g)$, which we call the \emph{%
coupling-deformed} (CD) observable hereafter.
We can get the weak-value numerator if the selection of $\hat{s}%
(g) $ makes the corresponding $Q$ matrix, $Q[g,\hat{s}(g)]$,
satisfy the relation
\begin{equation}
Q[g,\hat{s}(g)]=\eta Q(0,\hat{s})  \label{idea}
\end{equation}%
where $\eta $ is a proportionality constant, which may or may not depend
on $g$. Too see it, from Eqs. (\ref{key}) and (\ref{idea}) we have
\begin{equation}
\begin{aligned} \langle\Pi_f\otimes\hat{s}(g)\rangle_g =\eta
\langle\Pi_f\otimes\hat{s}\rangle_{g\rightarrow 0}. \end{aligned}
\end{equation}%
Then, as an alternative to Eq. (\ref{real-ima-weak}), if the corresponding $%
\hat{q}(g)$ and $\hat{p}(g)$ exist we have
\begin{equation}
\begin{aligned} &\mathfrak{Re}\,[\mathrm{tr}(\Pi_f\hat{A}\rho_{in})]=
\frac{i}{\eta
g\langle\phi_0|[\hat{q},\hat{p}]_{-}|\phi_0\rangle}\langle\Pi_f\otimes%
\hat{q}(g)\rangle_{g},\\
&\mathfrak{Im}\,[\mathrm{tr}(\Pi_f\hat{A}\rho_{in})]= \frac{1}{\eta
g\langle\phi_0|[\hat{p},\hat{p}]_{+}|\phi_0\rangle}\langle\Pi_f\otimes%
\hat{p}(g)\rangle_{g}. \end{aligned}  \label{reimweakvalues}
\end{equation}%
Therefore, we can obtain the weak-value information exactly in strong
measurements of any strength $g$, via reading the CD observables $\hat{q}(g)$
and $\hat{p}(g)$.

Then how can we identify the desired CD observable $\hat{s}(g)$ ($\hat{s}=\hat{q}$
or $\hat{p}$)? At first, one can choose an arbitrary orthonormal basis $\{|\tilde{u%
}\rangle \}$ of the space spanned by $\{|\phi _{m}(g)\rangle \}$, and define
the matrix $S(g)$ with elements given by $S_{um}=\langle \tilde{u}|\phi
_{m}(g)\rangle $. Suppose $\{|\phi _{m}(g)\rangle \}$ are linearly
independent (we will discuss the other case later); the matrix $S(g)$ will have
a well-defined inversion $S(g)^{-1}$. One can define a matrix $\tilde{Q}(g,%
\hat{s})$ by
\begin{equation}
\tilde{Q}(g,\hat{s})=\eta \;[S^{\dagger }(g)]^{-1}Q(0,\hat{s})S(g)^{-1},
\label{sg}
\end{equation}%
where $Q(0,\hat{s})$ is given by (\ref{order-1}). Then, the CD observable
can be chosen as
\begin{equation}
\hat{s}(g)=g\sum_{u,v}(\tilde{Q}(g,\hat{s}))_{uv}|\tilde{u}\rangle \langle
\tilde{v}|.  \label{determine-CD}
\end{equation}%
It is straightforward to show that the choice of CD observable according to (%
\ref{determine-CD}) ensures the requirement (\ref{idea}), and therefore also
ensures (\ref{reimweakvalues}). So one can choose $\hat{q}(g)$ [or $\hat{p}%
(g)$] according to (\ref{determine-CD}), and then retrieve the real (or
imaginary) part of the weak-value information $\mathrm{tr}(\Pi _{f}\hat{A}%
\rho _{in})$ from (\ref{reimweakvalues}) by measuring the expectation value
of $\Pi _{f}\otimes \hat{q}(g)$, or $\Pi _{f}\otimes \hat{p}(g)$, with any
strength $g$.

Here are some remarks. Equation (\ref{determine-CD}) fixes the effective
parts of $\hat{s}(g)$; one can add irrelevant terms living outside of the
space spanned by $\{|\phi _{m}(g)\rangle \}$. The proportionality constant $%
\eta $ in Eq. (\ref{sg}) is irrelevant to the statistics of the outcomes of
measuring $\hat{s}(g)$ and thus can be fixed by the convenience. The choice of $%
\hat{s}(g)$ is independent of the initial system state $\rho _{in}$ and thus
can be accomplished beforehand in the step of pointer calibration.

\subsection{Example: Weak-value tomography}
Now we show how to use our method of CD observables to extract weak-value information.
Let us revisit the experiment of Lundeen \emph{et
al.} \cite{wave function}, where a photon's spatial wave function is coupled
to polarization (pointer) via $U=\exp (-ig|x\rangle \langle x|\otimes \sigma
_{x})$ (we have $\hat{p}\rightarrow \sigma _{x}$). Since $\hat{A}%
=|x\rangle \langle x|$ is a rank-1 projector with $a_{0}=0$ and $a_{1}=1$,
there are only two relevant pointer states, $|\phi _{0}\rangle =|0\rangle $
and $|\phi _{1}(g)\rangle =\exp (-ig\sigma _{x})|\phi _{0}\rangle =\cos
(g)|0\rangle -i\sin (g)|1\rangle $ ($\sigma _{z}|1\rangle =-|1\rangle $).
Thus the $Q$ matrices will be 2 dimensional. In order to retrieve the real
part of the weak value, one has $\hat{q}\rightarrow \sigma _{y}$. According
to Eq. (\ref{order-1}),
\begin{equation}
Q(0,\hat{q})=-\left(
\begin{array}{cc}
0 & 1 \\
1 & 2%
\end{array}%
\right) .  \label{wave}
\end{equation}%
Now a convenient choice of the orthogonal basis of the space spanned by $%
\{|\phi _{0}\rangle ,|\phi _{1}(g)\rangle \}$ is $\{|0\rangle ,|1\rangle \}$%
, with which the matrix $S(g)$ introduced above can be written as
\begin{equation}
S(g)=-\left(
\begin{array}{cc}
1 & \cos (g) \\
0 & -i\sin (g)%
\end{array}%
\right) .  \label{S-matrix}
\end{equation}%
Then the matrix $\tilde{Q}(g,\hat{q})$ given by
Eq. (\ref{sg}) is
\begin{equation}
\tilde{Q}(g,\hat{q})=\frac{\eta }{\sin (g)}\left(
\begin{array}{cc}
0 & -i \\
i & -2\tan (\frac{g}{2})%
\end{array}%
\right) .  \label{tilde-Q}
\end{equation}%
Fixing $\eta =\sin (g)/g$ for convenience, from Eq. (\ref{determine-CD}) we
obtain the CD observable
\begin{equation}
\hat{q}(g)=\sigma _{y}-\tan (\frac{g}{2})(I-\sigma _{z}),\;\;\;g\in \lbrack
0,\frac{\pi }{2}].  \label{real-q}
\end{equation}%
According to (\ref{reimweakvalues}), we have
\begin{equation}
\mathfrak{Re}\,[\mathrm{tr}(\Pi _{f}\hat{A}\rho _{in})]=\frac{-1}{2\sin (g)}%
\langle \Pi _{f}\otimes \hat{q}(g)\rangle _{g}.
\end{equation}%
The case of a strong limit achieved at $g=\frac{1}{2}\pi $
is also studied in Ref. \cite{strong-chi}, and cases of arbitrary
strength in Ref. \cite{arxiv-strong}. Their methods are
based on Eq. (\ref{expansion}). An analytical result is possible
because Lundeen's setup involves only a simple $\hat{A}$ (projector) and
simple pointer (qubit), while our result origins
from a more general approach with CD observable.

\subsection{Resolution of the drawbacks}

Here we explain how the shortcomings of the previous
weak-value tomography listed in Ref. \cite{simulation} are overcome by our approach
of CD observables.
A more elaborate study is reported in an independent work with other colleagues \cite{zhu-num}.

When the
tiny but finite coupling strength is fixed, the pointer reading will
deviate from the $g A_w$ when $|\psi_{in}\rangle$ tends to be orthogonal to
$|\psi_f\rangle$ \cite{wu-li,pang}. Our method will not face such possible failure,
because the formalism
is exact and universally valid.

The formalism of the weak-measurement scheme is exact only in the weak limit,
while the real experiments need a finite interaction strength. 
This discrepancy causes the
bias, the deviation between the expectation of the estimator (the reconstructed states) 
and the real value of the estimated state. 
In weak-value tomography,
this bias is confirmed to be
very robust \cite{simulation}, i.e., hard to be removed, 
while here it automatically disappears
(we presume the perfect implementation) because of the exactness of our formalism in the whole range of coupling strength.

The most important improvement is the reduction of quantum noise. With
our formalism, weak-value tomography can be implemented using stronger
interactions. The efficiency of information extraction will be superior
to a significant extent.

To verify the argument, we performed a numerical simulation of
three tomography methods, the traditional weak-measurement schemes, the
standard tomography, and our method with CD pointer observables. We
randomly select a qubit state, simulate N times of quantum
measurements, and then calculate the trace distance between the real state
and the reconstructed state.

\begin{figure}[bt]
\includegraphics[width=0.44\textwidth]{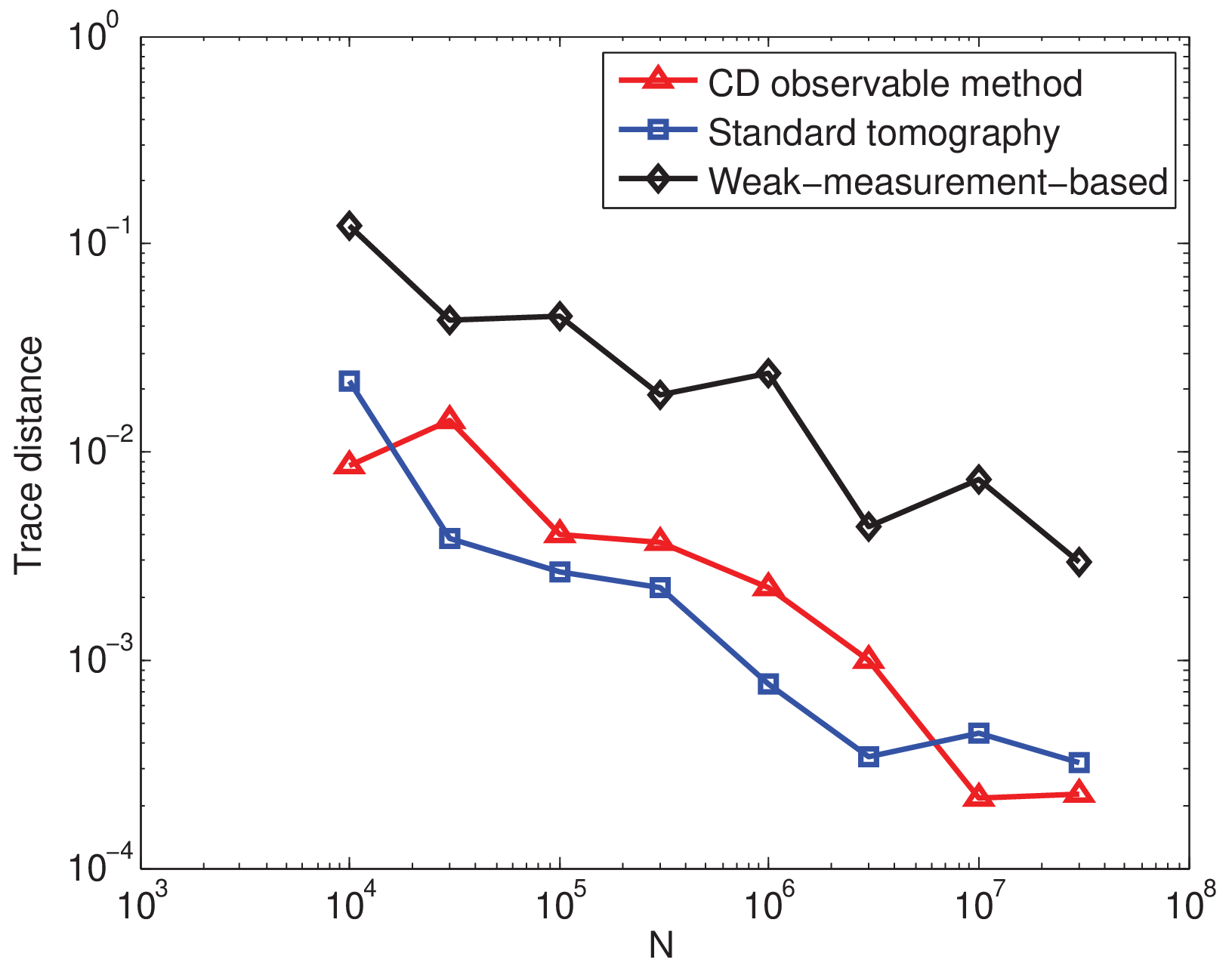}
\caption{Trace distance (between the real and the reconstructed states)
against the times of repetitions (sample size, N). The weak-value tomography
based on weak measurements is simulated with the strength $g=0.1$,
our method with the CD observables is simulated with the strength $g=\protect%
\pi/2$.}
\label{fig-2}
\end{figure}

The numerical result is illustrated in Fig. \ref{fig-2}, which gives curves
of reconstruction error (trace distance) against the times of repetition
(sample size). It shows that, to reach the same level of precision, the
method based on weak measurements needs many more samples than the other
two, while the sample size used in our method with the CD observables is
comparable with that of the standard tomography. Thus,
the problem of resource consumption is cured.
A more thorough comparison between the three is reported in Ref. \cite{zhu-num}.

\subsection{The $g$-invariant CD observables}
There are situations when the CD
observable $\hat{s}(g)$ is independent of $g$. Such \emph{g-invariant CD
observables} could simplify the experimental implementation.
The $g$-invariant observables will also be
necessary for applications when the coupling constant is either unknown or
uncertain with inevitable errors. Below we give examples of $g$-invariant
observables that are easy to measure.

\subsubsection{$\hat{A}$ is a projector}
We consider again the experiment of Lundeen \emph{et al.} \cite{wave
function} in which one can obtain the imaginary part of the weak value by
measuring $\hat{p}\rightarrow \sigma _{x}$. We have
\begin{equation}
Q(g,\sigma _{x})=\frac{\sin (g)}{g}\left(
\begin{array}{cc}
0 & -i \\
i & 0%
\end{array}%
\right) =\frac{\sin (g)}{g}Q(0,\sigma _{x}).
\end{equation}%
Thus Eq. (\ref{idea}) is fulfilled, and $\sigma _{x}$ is certainly a \emph{g}%
-invariant CD observable in this case. If the pointer is replaced by a
continuous-variable system and initialized in a Gaussian state with standard
deviation $\Delta $,
\begin{equation}
\langle q|\phi _{0}\rangle =\frac{1}{(2\pi \Delta )^{1/4}}\exp (-\frac{q^{2}%
}{4\Delta ^{2}}),
\end{equation}%
which is very common in reported experiments on weak measurements \cite%
{review}, we will have ($\hat{p}=-i\frac{\partial }{\partial q}$)
\begin{equation}
Q(g,\hat{p})=\frac{1}{4\Delta ^{2}}e^{-\frac{g^{2}}{8\Delta ^{2}}}\left(
\begin{array}{cc}
0 & -i \\
i & 0%
\end{array}%
\right) =e^{-\frac{g^{2}}{8\Delta ^{2}}}Q(0,\hat{p}).
\end{equation}%
The CD observable is found to be the original observable $\hat{p}$, which is
of course $g$-invariant.

\subsubsection{$\hat{A}$ is effectively a Pauli operator}
Another case of 2-dimensional $Q$ matrix is
when $\hat{A}$ has only two nonzero eigenvalues while $|\psi_f\rangle$
resides in the support set of $\hat{A}$.

Suppose that $\hat{A}=a_1\Pi_1+a_2\Pi_2$ where
$a_{1(2)}\neq 0$ and $\Pi_{1(2)}$ is the projector onto the eigenspace.
Then the formalism of the system with variables combination
$(g,\hat{A},|\phi_0\rangle)$ is equivalent to that of another system where
the experimental variables are $(g_{eff},\hat{A}_{eff},|\phi_0'\rangle)$:
\begin{equation}
\begin{aligned}
 g_{eff} &=\frac{a_1-a_2}{2}g; \qquad
 \hat{A}_{eff} =\Pi_1-\Pi_2; \\
 |\phi_0 '\rangle & =\exp(-ig\frac{a_1+a_2}{2}\hat{p})|\phi_0\rangle.
\end{aligned}
\end{equation}
The equivalence is that, the elements of $Q$-matrices can be calculated
with the pointer states given by
\begin{equation}
|\phi_{\pm}\rangle=\exp(\pm ig_{eff}\hat{p})|\phi^{'}_0\rangle.
\end{equation}
Thus we just need to consider operators in the form of $\hat{A}_{eff}$,
a generalization of Pauli operator $\sigma_z$.
But note that if $(\Pi_1+\Pi_2)|\psi_f\rangle\neq |\psi_f\rangle$,
$|\phi_0 '\rangle$ should be included so that $Q$ matrices become three dimensional.

The investigated two-dimensional $Q$ matrices are summarized in Tab. \ref{tab-1},
where those support $g$-invariant pointer observables are marked.
Situations when $\hat{A}$ is the generalized Pauli
matrix are discussed in Ref. \cite{xiang}.
In the three cases marked in Tab. \ref{tab-1},
if the pointer is played by a qubit in state $|0\rangle$, then the pointer observables
can be chosen as Pauli operators $\hat{p}\rightarrow \sigma_x$
and $\hat{q}\rightarrow \sigma_y$,
while for the continuous-variable
pointers and the initial state $|\phi _{0}\rangle $ satisfying some conditions
to be given below, the CD observables will be g-invariant.

Suppose $\hat{q}$ is the position operator and $\hat{p}$ is the momentum
operator. Particularly, in the right column of Table \ref{tab-1}, the
diagonal elements of the $Q$ matrices (for the imaginary parts of the weak-value
numerator) are derived as
\begin{equation}
\langle\phi_0|e^{iga_m\hat{p}}\hat{p}e^{-iga_m\hat{p}}|\phi_0\rangle=\langle%
\phi_0|\hat{p}|\phi_0\rangle=0,
\end{equation}
provided that the
initial reading of the pointer is zero.
The two off-diagonal terms must be complex conjugate since $Q$
matrices are Hermitian by definition. Therefore, the
desired $Q$ matrices are ensured if
\begin{equation}
\langle \phi _{0}|\hat{p}\exp (-ig(a_{n}-a_{m})\hat{p})|\phi _{0}\rangle
\end{equation}%
is a pure imaginary number. It is sufficient to require the wave function of $%
|\phi _{0}\rangle $ in momentum representation, i.e., $\phi _{0}(p)$ to have
even parity.

\begin{table}
    \centering
   \begin{tabular}{c m{0.5cm} m{2cm} m{1.8cm}}
   \toprule
    $\hat{A}$  & \qquad & $ Q(0,\hat{q}) $  &   $Q(0,\hat{p})$\\
   \midrule
    $\Pi$  &  \qquad & -$\begin{pmatrix}  0 & 1\\ 1 & 2 \end{pmatrix}$
                    & $\begin{pmatrix} 0 & -i\\ i & 0 \end{pmatrix}_\surd$\\
    $\Pi_1-\Pi_2$  & \qquad & 2$\begin{pmatrix} 1 & 0\\ 0 & -1 \end{pmatrix}_\surd $
                   &  2$\begin{pmatrix} 0 & -i\\ i & 0 \end{pmatrix}_\surd$\\
    \bottomrule
  \end{tabular}
  \caption{The 2-dimensional $Q$matrices, $\hat{q}$ and $\hat{p}$
  are the pointer observables for the real and imaginary parts of weak-value information,
  respectively. In cases marked with ``$\surd$'', the CD observables are $g$-invariant
  when the pointers are initialized in proper states.}
  \label{tab-1}
\end{table}

For the g-invariant CD observable in the left column, being a continuous
variable state, $|\phi _{0}\rangle $ should satisfy the condition that
\begin{equation}
\langle \phi _{0}|[\hat{q},e^{i2g\hat{p}}]_{+}|\phi _{0}\rangle =0.
\end{equation}%
It degenerates to the condition $\langle \phi _{0}|[%
\hat{q},\hat{p}]_{+}|\phi _{0}\rangle =0$ in the weak limit.

\section{Methods against poor pointers}\label{section poor}
Our general method with CD
observables requires the linear independence of $\{|\phi_m(g)\rangle\}$. But
if it cannot be fulfilled due to a pointer with limited dimensions of Hilbert
space, or if the CD observables are either hard to calculate or hard to
measure in practice, do we have methods to circumvent these troubles?

One strategy is to measure another observable, say $\tilde{A}$, which supports
2-dimensional $Q$ matrices, and of which the weak value has the same information of $A_w$.
One construction is like this: define
\begin{equation}
|\psi _{A}\rangle =\mathcal{N}\sum_{m}a_{m}\langle a_{m}|\psi _{f}\rangle
|a_{m}\rangle ,
\end{equation}%
where $\mathcal{N}=1/\sqrt{\langle\psi|\hat{A}^2|\psi_f\rangle}$
is the normalization factor. Then we have
\begin{equation}
\Pi _{f}\hat{A}\propto \Pi _{f}|\psi _{A}\rangle \langle \psi _{A}|.
\end{equation}%
where the proportionality factor is
$\frac{\langle\psi_f|\hat{A}^2|\psi_f\rangle}{\langle\psi|\hat{A}|\psi_f\rangle}$.
It implies that $\tilde{A}$ can be the rank-1 projector, $|\psi
_{A}\rangle \langle \psi _{A}|$, provided that $\langle \psi _{f}|\psi
_{A}\rangle\neq 0$.
Otherwise $\Pi _{f}|\psi _{A}\rangle \langle \psi _{A}|$ is
trivially zero. In this case, $\tilde{A}$ can be selected as the generalized Pauli matrix $%
|\psi _{A}\rangle \langle \psi _{f}|+|\psi _{f}\rangle \langle \psi _{A}|$
which allows 2-dimensional $Q$ matrices.
Then even a poor pointer with only 2-dimensional Hilbert space can be used
to extract weak-value information.

Another strategy is to modify the postselection.
Consider the following equality
\begin{equation}
\mathrm{tr}(\Pi _{f}\hat{A}\rho _{in})=[\mathrm{tr}(\hat{A}\Pi _{f}\rho
_{in})]^{\ast }.  \label{trans}
\end{equation}%
That is, we can use $\tilde{A}=\Pi_f$ and replace the original postselection
with a projective measurement of $\hat{A}$, i.e., project the system onto $\{|a_m\rangle\}_m$
and multiply the pointer readings by the corresponding $a_m$ (when the system is projected
onto $|a_m\rangle$).
Then Eq. (\ref{trans}) shows that the resulted value is just
the complex conjugation of the original weak value.
We remark here that since $\tilde{A}$ is a projector here, the $Q$ matrices are
2 dimensional. Meanwhile, the CD observables are irrelevant
to the postselected state. Therefore, only a fixed setting of pointer system is needed, although the postselection is performed
onto a set of orthogonal states.

\section{Discussion and Conclusions} \label{section discussion}

We would like to discuss more the experimental realizations. 
Given a weak-measurement scheme, our main result implies that, the
measurement can be strengthened if the CD observables can be read on the pointer.
That means the feasibility of our method relies on the
presumption that the pointer can be operated conveniently.
Otherwise, we have to measure $\tilde{A}$ instead of the original $\hat{A}$,
as discussed in Sec. \ref{section poor},
while in the important example of weak-value tomography, 
our method is very easy to realize within Lundeen's setup \cite{wave function}, 
where the pointer is played by the polarization freedom.
Operators of any direction (in the Bloch sphere)
can be simply measured.
Meanwhile, the coupling between spatial wave function and
polarization (the pointer) is realized with a rectangular sliver of a
half-wave plate. The coupling strength is determined by the angle shift on
the photons' polarization. To implement stronger measurements, we just need
to tune larger this angle. Therefore, everything
can be done easily.

Our result also has other implications. For example, the accessibility of
weak-value information is often attributed to the negligible disturbance
caused by weak measurements in the literature; see, e.g., Refs. \cite{wave
function,trajectory}. However, our results show that such interpretation is
unnecessary. We hope this work could stimulate more sparks on theories and
applications of ``weak'' measurements.

To summarize, we have developed a
nonperturbative approach to retrieve weak-value information in measurements
with post-selections. Here the system-pointer coupling strength can be of
any finite value, not necessarily small. To retain the original form of the
weak-value, we can slightly modify the current weak-measurement scheme, and
read the CD observables on the pointers instead. We also studied situations
when such a modification is unnecessary, namely, the CD observables are $g$%
-invariant. This is meaningful for simplifying the experiments. Thus, while
keeping the advantages of current weak-measurement and weak-value motivated
applications, our method eliminates main problems therein, such as
inefficiency, bias, and problematic
universality in the current weak-measurement tomography scheme, without
introducing much complexity in experimental implementations.

\acknowledgments
We thank Xuanmin Zhu, Yi-Zheng Zhen, and the
anonymous referees for valuable suggestions. This work was supported by the
National Natural Science Foundation of China (Grants No. 11475084 and No.
61125502), the National Fundamental Research Program of China (Grant No.
2011CB921300), and the CAS.

\end{document}